\documentclass[10pt,conference]{IEEEtran}
\IEEEoverridecommandlockouts
\usepackage[table]{xcolor}  
\usepackage{array}       
\usepackage{booktabs}   

\usepackage{algorithm}
\usepackage{algorithmic}
\usepackage{multirow}
\usepackage{graphicx}
\usepackage{listings}
\usepackage{xcolor}
\usepackage{hyperref}
\usepackage{hyperxmp}
\usepackage{amsmath}
\usepackage[skins,breakable]{tcolorbox}
\usepackage{cite}
\usepackage{balance}

\def\BibTeX{{\rm B\kern-.05em{\sc i\kern-.025em b}\kern-.08em
    T\kern-.1667em\lower.7ex\hbox{E}\kern-.125emX}}

\lstdefinestyle{prompt}{
    basicstyle=\ttfamily\footnotesize,
    backgroundcolor=\color{gray!10},
    frame=single,
    breaklines=true,
    keepspaces=true,
    columns=fullflexible,
    captionpos=b
}

\newcommand{\find}[1]{%
  \begin{tcolorbox}[
    leftrule=1mm, toprule=0mm, bottomrule=0mm,
    breakable,
    left=1pt,  right=2pt,
    top=1pt,   bottom=1pt,   
    boxsep=1pt,              
    fontupper=\small
    ]%
    \em #1%
  \end{tcolorbox}}

\newcommand{\rev}[1] {{\textcolor{black}{#1}}}

\newcommand{\sectopic}[1]{\vspace{0.2em}\par\noindent{\textit{\bfseries #1}}}

\newcommand{\projName}{\textit{UniLearn}}
\newcommand{\company}{\textit{eSolutions}}
\newcommand{\approachName}{\textit{RAGcceptance\_M2RE}}

\begin{document}

\title{Multi-Modal Requirements Data-based Acceptance Criteria Generation using LLMs}

\author{
\IEEEauthorblockN{%
Fanyu Wang\IEEEauthorrefmark{1}, Chetan Arora\IEEEauthorrefmark{1}, Yonghui Liu\IEEEauthorrefmark{1}
Kaicheng Huang\IEEEauthorrefmark{1}, \\Chakkrit Tantithamthavorn\IEEEauthorrefmark{1}, Aldeida Aleti\IEEEauthorrefmark{1}
Dishan Sambathkumar\IEEEauthorrefmark{2}, David Lo\IEEEauthorrefmark{3}}
\IEEEauthorblockA{\IEEEauthorrefmark{1}\textit{Faculty of Information Technology}, Monash University, Clayton, VIC, Australia\\
\texttt{\{fanyu.wang, chetan.arora, yonghui.liu, khua0042, chakkrit, aldeida.aleti\}@monash.edu}}
\IEEEauthorblockA{\IEEEauthorrefmark{2}eSolutions, Monash University, Clayton, VIC, Australia\\
\texttt{dishan.sambathkumar@monash.edu}}
\IEEEauthorblockA{\IEEEauthorrefmark{3}\textit{School of Computing and Information Systems}, Singapore Management University, Singapore\\
\texttt{davidlo@smu.edu.sg}}
}

\maketitle

\begin{abstract}
Acceptance criteria (ACs) play a critical role in software development by clearly defining the conditions under which a software feature satisfies stakeholder expectations. However, manually creating accurate, comprehensive, and unambiguous acceptance criteria is challenging, particularly in user interface-intensive applications, due to the reliance on domain-specific knowledge and visual context that is not always captured by textual requirements alone. To address these challenges, we propose \approachName, a novel approach that leverages Retrieval-Augmented Generation (RAG) to generate acceptance criteria from multi-modal requirements data, including both textual documentation and visual UI information. We systematically evaluated our approach in an industrial case study involving an education-focused software system used by approximately 100,000 users. The results indicate that integrating multi-modal information significantly enhances the relevance, correctness, and comprehensibility of the generated ACs. Moreover, practitioner evaluations confirm that our approach effectively reduces manual effort, captures nuanced stakeholder intent, and provides valuable criteria that domain experts may overlook, demonstrating practical utility and significant potential for industry adoption. This research underscores the potential of multi-modal RAG techniques in streamlining software validation processes and improving development efficiency. We also make our implementation and a dataset available.

\end{abstract}


\section{Introduction}
\label{sec:introduction}

Acceptance testing is a crucial validation activity within the software development lifecycle, ensuring that a software system not only meets functional requirements but also aligns with user expectations~\cite{fischbach2023automatic,clark2021test}. Distinguished from other testing approaches, acceptance testing focuses explicitly on compliance with the requirements specification~\cite{perala2021review}, establishing a strong connection with the requirements engineering (RE) stage. 
At the core of \emph{acceptance testing} are acceptance criteria, which serve as the principal test specifications for acceptance testing. Acceptance criteria (ACs)\footnote{We abbreviate multiple criteria as ACs and a single criterion as AC.} are defined as ``the conditions that a system must meet to fulfil a user story and ultimately be accepted by the user''~\cite{fischbach2020makes}. This definition highlights two key requirements: (1) acceptance criteria must be well-aligned with the requirements, and (2) they must effectively capture the user’s intent. Distinguishing from other test artifacts (e.g., unit or integration tests), which typically verify concrete implementation in primary~\cite{rwemalika2023smells}, ACs are not only designed to reflect high-level user or business objectives, domain-specific nuances clearly, but also often abstract concepts conveying user interactions and system behaviors. 

In professional software development environments, ACs are predominantly documented in structural natural language~(NL). The Gherkin format has been well adopted as the typical format of ACs, following the ``GIVEN'', ``WHEN'', and ``THEN'' pattern~\cite{dos2018automated,karpurapu2024comprehensive}. 
The manual specification of ACs requires a comprehensive understanding of both the requirements and the context, as well as a significant time investment to ensure accuracy and clarity. This process becomes even more challenging due to ambiguities, inadequate details in user stories, and the evolving nature of software specifications. 


Automating quality assurance artefact generation, like ACs, is an important task in the industry~\cite{wang2025requirements}. However, the inherent challenges in manual creation persist in automated approaches, necessitating solutions that ensure accurate intent understanding and requirements alignment.
Large Language Models (LLMs) have demonstrated remarkable capabilities in natural language processing (NLP), and automating software engineering (SE) tasks~\cite{fan2023large, nguyen2023generative}. Despite their advanced capabilities, concerns remain regarding their ability to acquire domain-specific knowledge and fully grasp nuanced user intent~\cite{xue2024domain,gu2025effectiveness,welz2024enhancing}.
In the context of ACs, an unguided LLM might inadvertently `hallucinate' and invent conditions not specified by stakeholders or omit essential constraints that were specified. For example, given a vaguely worded requirement, an LLM might fill in gaps with its own assumptions, producing criteria that appear reasonable but do not actually reflect the true intent of the requirement. This undermines trust in automation: stakeholders and developers cannot rely on generated criteria if there is a risk of false conditions leading the project astray. Ensuring that every generated AC is faithful to the source requirements (and nothing more) is therefore paramount when introducing automation.

To address these challenges, we propose \approachName, which stands for Retrieval Augmented Generation based acceptance criteria generation from Multi-Modal REquirements information (\textsf{reads RAGcceptance motor}). Our approach combines the generative power of LLMs with a retrieval mechanism to automate ACs generation in a more reliable way. By integrating Retrieval-Augmented Generation (RAG)~\cite{lewis2020retrieval} with multi-modal requirements representations, this method grounds the LLM’s output in authentic project data. Rather than relying solely on the LLM, \approachName~automatically retrieves pertinent context from existing domain artifacts such as project background information \rev{(the interaction logics, stakeholders' description, or system objectives)} and system UI, providing context to the LLM as additional input. This additional context ensures that each generated criterion is based on authentic project data rather than assumptions. In our paper, we leverage multi-modal RAG, i.e., in addition to using RAG for textual artifacts (which have been previously used in RE research~\cite{arora2024generating}), we leverage visual RAG models for incorporating additional information captured from the system's UI. We posit that integrating visual elements with the textual domain information enhances the accuracy and completeness of the generated ACs, resulting in outputs closely aligned with stakeholder expectations.

\subsection{Industry Context}
We evaluate \approachName~in an industrial study conducted with \company, which is the central IT partner for Monash, and as an individual entity comprises over 500 ongoing staff and an additional 200 contractors, casuals, and affiliates. It delivers end-to-end IT services across the university—from installing software on student devices to developing and maintaining the world’s largest online exam platform and the infrastructure that underpins it. Their projects encompass course administration, IT service provision, building software products, and meeting all infrastructure, software, and cybersecurity needs. We conducted our evaluation on \projName~product data~(some details are anonymized due to proprietary nature), which is a representative product in their series of customised products. \projName~is a university learning management system with $\sim$100,000 current users.

\begin{figure}
    \centering
    \includegraphics[width=0.88\linewidth]{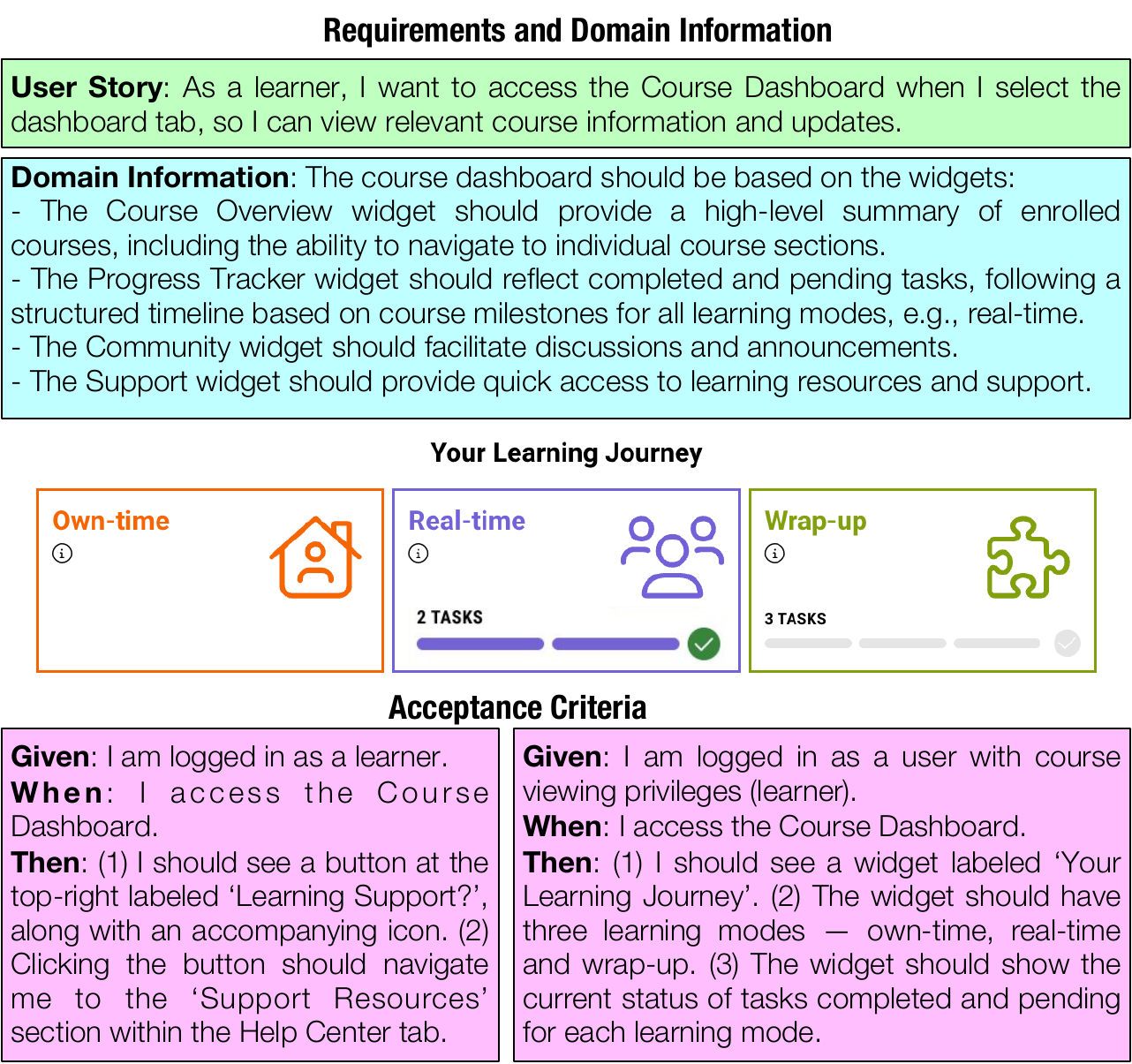}
    \vspace*{-0.5em}
    \caption{Example User Stories, Domain Information, \rev{System UI}, and Acceptance Criteria from \projName~project at \company~organisation.}
    \label{fig:running_example}
    \vspace*{-2em}
\end{figure}

\subsection{Working Example and Motivation}
Fig.~\ref{fig:running_example} shows an example user story, multi-modal domain information, and ACs\footnote{\rev{This example is taken \emph{verbatim}—with all their minor formatting inconsistencies and non-standard labels to show the truly ``noisy'' industry inputs}.}. This example has been slightly adapted from the original dataset from \projName. The user story in the example is related to the customisation for learner groups. 
The ACs in the example are also closely related to the original ACs specified by analysts. Having said that, the figure shows ACs that are composite, i.e., each criterion contains multiple sub-criteria. However, our approach generates atomic criteria, based on the best practices~\cite{cohn2004user} and recommendations by the \company~quality assurance (QA) team.

The user story in the example only mentions ``relevant course information and updates'' -- which is insufficient for generating accurate ACs. Analysts usually rely on their domain knowledge to write complete and accurate ACs, which is not captured in user stories. The first specified AC mentions ``learning support'' which is captured from the textual domain information from the project background document, which usually has much more information and hence contributes to writing ACs. The second AC relies on both the textual and visual information (see a real UI screenshot from \projName) of learning modes of own-time, real-time and wrap-up, and the corresponding tasks assigned to learners in a given course. As evident in Figure~\ref{fig:running_example}, background information is required to supplement the information captured in user stories. Specifying ACs manually requires deep domain expertise, which is often limited to a subset of project analysts. An approach like \approachName~is targeted at supporting analysts who might be new to projects or save effort on specifying ACs even for experienced analysts.

\subsection{Contributions}
This paper introduces \approachName, an innovative approach for automating the generation of ACs from multi-modal requirements information. \rev{We release our implementation and testset~\cite{ourRepo}.\footnote{\rev{The testset is a small anonymized subset from \projName. We remove any confidential/proprietary information from the subset.}}} Our key contributions are:

$\bullet$ \textbf{Multi-Modal Retrieval-Augmented Generation (RAG):} We propose an integrated framework that combines textual and visual RAG techniques. This ensures comprehensive and context-aware AC generation by retrieving and utilizing relevant textual and visual project artefacts. To the best of our knowledge, our study is among the first to explore and validate multi-modal RAG for requirements-driven automated quality assurance in software engineering, establishing a robust foundation for future research and practical applications.

$\bullet$ \textbf{Reward-based Refinement Process}: We incorporate two distinct reward models, commonly used in the AI domain, to iteratively refine generated ACs, significantly improving their correctness, clarity, and alignment with user stories.

$\bullet$ \textbf{Empirical Evaluation in an Industrial Context}: Our method is rigorously evaluated in an industrial context involving data from \projName~product from \company. Industrial evaluations are rather scarce in the RE research domain, and our evaluation demonstrates substantial performance improvements and validates practical applicability.

$\bullet$ \textbf{Expert Validation}: We gathered detailed feedback from experienced practitioners, providing practical insights into the quality, utility, and effectiveness of our generated ACs. Experts confirmed the significant potential for time savings and highlighted instances where our method captured essential criteria not explicitly stated in user stories.

\sectopic{Paper Structure.} Section~\ref{sec:background} provides background on relevant concepts and related work. Section~\ref{sec:approach} outlines our technical approach. Section~\ref{sec:Evaluation} details our industry evaluation. Section~\ref{sec:threats} discusses threats to validity, and Section~\ref{sec:conclusion} concludes.

\section{Background}
\label{sec:background}
This section provides two key concepts: retrieval augmented generation (RAG) and reward models used in our approach. We further position our work against the related work.

\subsection{Retrieval-Augmented Generation (RAG)}
RAG is a recent significant advancement in the NLP domain. First formalized by Lewis et al.~\cite{lewis2020retrieval}, it is defined as the process of integrating the external knowledge sources in the generation process with the help of pre-trained LLMs~\cite{yu2024evaluation, gupta2024comprehensive}. The typical components in the RAG pipeline consist of three stages, including indexing, retrieval, and generation~\cite{du2024vul}. Given a query, the RAG module first searches through the \textit{indexed} structure of external knowledge to \textit{retrieve} the most relevant target. The retrieved knowledge is then used in the \textit{generation} stage, where it is fed along with a prompt to an LLM to perform the generation task, e.g., generating ACs in our case. RAG models, compared to LLMs for generation alone, help alleviate hallucination issues, particularly when project-specific context is unavailable as pre-training for LLMs~\cite{arora2024generating,yu2024evaluation}. RAG models can have multi-modalities~\cite{10.1145/3637528.3671470, hu2023reveal}. In our work, we use both textual and visual RAG models.


\sectopic{Textual RAG (T-RAG)} is the most common scenario in practice, serving as the foundational approach for RAG models. Traditional information retrieval techniques, such as TF-IDF~\cite{ramos2003using} and BM25~\cite{trotman2014improvements}, are often utilized, particularly in recommendation systems, to identify relevant documents from large textual datasets efficiently. With advancements in transformer networks, discrete words can be converted into integrated embedding vectors, which further support correlation computation between queries and external documents, using cosine similarity~\cite {juvekar2024cos,csakar2025maximizing}. Classic T-RAG models typically involve indexing documents through embeddings, such as SentBERT~\cite{reimers2019sentence}, which transforms textual data into dense vector representations that capture semantic meaning. These embeddings enable effective retrieval through semantic similarity metrics. For example, SentBERT embeddings are commonly used due to their effectiveness in capturing sentence-level semantics, making them highly suitable for T-RAG. Another variation is \emph{In-context textual RAG models}, which further refine the retrieval-generation interaction by explicitly embedding retrieved textual examples directly into the prompt provided to the LM. By including retrieved-context as explicit examples in the prompt, the model can generate outputs that closely match the exemplified style, structure, and content. In our work, we experiment with two different T-RAG alternatives -- SentBERT (as discussed above) and ICRALM (In-Context Retrieval-Augmented Language Models)~\cite{ram2023context}.

\sectopic{Visual RAG (V-RAG)} extends the traditional T-RAG approaches by incorporating visual information into the retrieval-generation pipeline. V-RAG models index and retrieve relevant visual artefacts, such as images, screenshots, or interface mock-ups, leveraging visual embedding techniques from advanced vision models, e.g., MuRAG~\cite{chen2022murag}. Given a textual query, V-RAG models retrieve visually similar artefacts by matching textual embeddings with visual embeddings in a shared semantic space. The retrieved visual context provides richer and more concrete grounding, significantly enhancing the accuracy, specificity, and context-awareness of generated outputs. Such visually informed generation is especially valuable in tasks requiring visual comprehension, such as generating UI-related acceptance criteria. V-RAG models are also based on multi-modal LLMs, e.g., RA-CM3~\cite{yasunaga2022retrieval}, which integrates a base multi-modal LLM in the framework to generate mixtures of text and images.


\subsection{Reward Model}
\label{sec:reward_model} A reward model helps evaluate how good or appropriate an output is from an AI model. Reward models originally gained prominence with systems like AlphaGo~\cite{silver2017mastering}, where a reward was used to evaluate the outcomes of various actions. In the context of LLMs, a reward model is a specialized mechanism designed to assess the quality and appropriateness of generated outputs~\cite{reber2024rate}. These models use a classifier architecture that assigns scores or labels to outputs, reflecting their relevance, accuracy, or preference alignment~\cite{liu2024rm}. Common approaches include probability-based reward functions~\cite{lee2017muse,huyuk2022inferring}, explicit class-based scoring with clearly interpretable scales (e.g., ratings 0--4)~\cite{eshuijs2024balancing,kim2024prometheus,wang2024helpsteer2}, and using LLMs themselves as judges, comparing candidate outputs directly to select the most appropriate one based on provided context~\cite{thakur2024judging,li2024llms}. Such clearly defined scoring systems make the reward model outputs interpretable, enabling practical assessment and ongoing model improvement. We use different reward models in our approach to iteratively improve the quality of the generated ACs. In our work, we have two levels of the rewarding process, where Prometheus~\cite{kim2024prometheus} serves as the higher level of reward model for overall scoring for the generated ACs based on five predefined quality levels: generative verifier~\cite{zhang2024generative} and UR3~\cite{yuan-etal-2024-improving} are two baselines for the lower level of reward process. The details of our reward processes are introduced in Section~\ref{sec:quality_assurance}.

\subsection{Related Work}
Requirements-driven automated quality assurance artifact generation remains underexplored in software engineering~\cite{mustafa2021automated}. A recent survey by Wang et al.~\cite{wang2025requirements} identifies 156 related studies published between 1990 and 2024. Notably, the survey highlights that NL requirements, due to their inherent ambiguity and incompleteness~\cite{dar2022reducing,dar2024gamify4lexamb}, receive relatively little attention, accounting for only 30\% of all studies~\cite{chittimalli2008regression,almohammad2017reqcap,verma2013representation}. Acceptance criteria play a crucial role in acceptance testing by providing a clear and objective framework to evaluate whether a product or system meets the specified requirements and standards~\cite{dos2024aat4irs,bjarnason2016multi,brockenbrough2025exploring}. Compared to other test artifacts, ACs exhibit a more structured format, typically following the ``GIVEN'', ``WHEN'' and ``THEN'' pattern~\cite{dos2018automated,karpurapu2024comprehensive}. This structure demands a deeper understanding of stakeholder expectations, more concrete test implementation, and greater clarity~\cite{fischbach2020makes}. The officially reported studies on this topic are very few, where Wang et al.~\cite{wang2025requirements} identify only one study that explicitly addresses the generation of acceptance criteria~\cite{guldali2011torc}. Aside from acceptance criteria, there are few studies on the generation of automated acceptance testing. Fischbach et al.~\cite{fischbach2023automatic} proposed CiRA, a conditional statements-based approach using NLP tools for acceptance test case creation. Wang et al.~\cite{wang2020automatic} introduced an automated acceptance testing generation approach from use case specifications using NLP pipelines, including POS tagger, tokenizer, and others. Straszak et al.~\cite{straszak2014automating} proposed an automated acceptance testing method with some RSL (Requirements Specification Language) parsing tool. AGUTER is a platform to design UATs (User Acceptance Testing) from scenarios using the task/method model.

Our framework introduces advanced NLP techniques, specifically multi-modal retrieval-based generation, to enable diverse domain documents that support the generation of acceptance criteria. The framework design addresses the ambiguity of NL requirements by utilizing advanced LLMs with reward processes and implements concrete acceptance criteria using a multi-modal RAG. However, as far as we know, it is the first work to enable multi-modal requirements-driven automated software testing.
\section{Approach}
\label{sec:approach}
\begin{figure}
    \centering
    \includegraphics[width=0.9\linewidth]{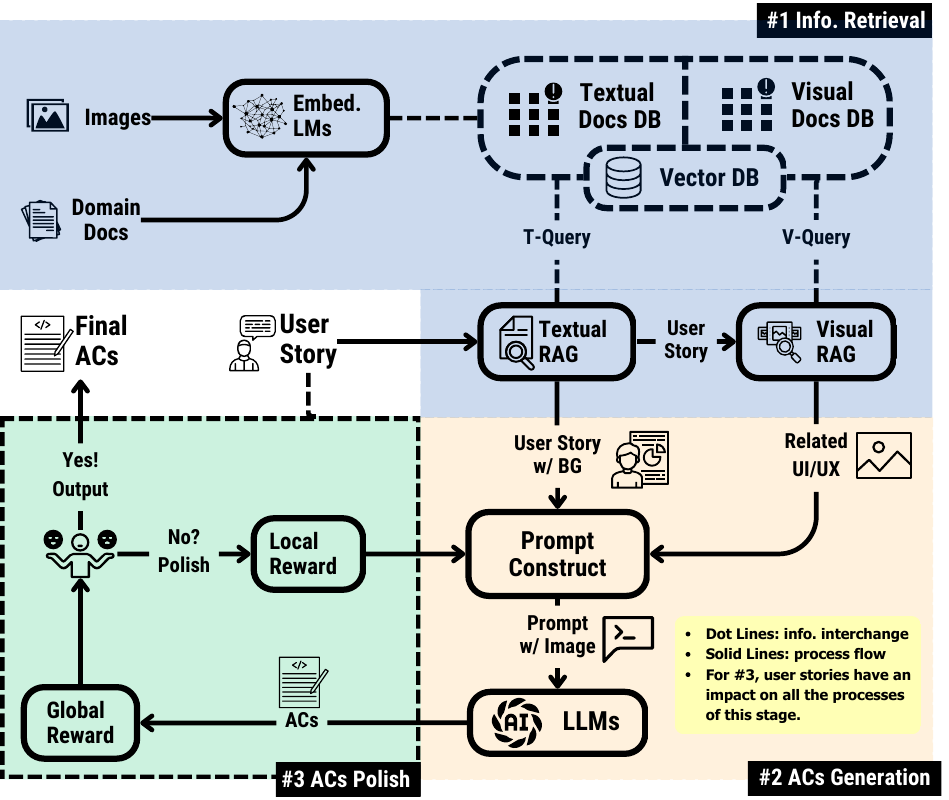}
    \vspace{-1em}
    \caption{Overview of \approachName}
    \label{fig:framework}
    \vspace{-2em}
\end{figure}

Fig.~\ref{fig:framework} shows an overview of \approachName~for generating ACs from NL requirements, specified as user stories. The inputs to our approach are user stories, with (optional) additional background information in the form of textual domain documents and UI in the form of screenshots. The output of our approach are acceptance criteria generated for the user story. Our approach consists of three stages: (1)~Domain Information Retrieval, (2)~Acceptance Criteria Generation, and (3) Reward-based Post-Processing.

\subsection{Domain Information Retrieval}
Given a user story, the first step in our approach is to retrieve relevant contextual domain information. User stories provide high-level requirements information, often rendering themselves insufficient for generating detailed and concrete acceptance criteria. To address this limitation, the information retrieval stage is introduced to leverage user stories as input to query relevant domain knowledge. Our approach supports textual information retrieval and enhances the retrieval process with image-based knowledge sources by integrating both textual and visual RAG methods, greatly enhancing the contextual information and the user story. 

\sectopic{Textual Information Retrieval.}~\label{sec:trag}
Given a user story in textual format, the textual RAG (T-RAG) model predicts a relevance score for each domain document based on its relatedness to the user story. After ranking the passages, the top $k$-ranked ones are selected for retrieval. One can use several measures for determining the similarity between the query (in our case, the user story) and the underlying domain documents for RAG models. We experiment with two alternative techniques, namely \emph{SentBERT}~\cite{reimers2019sentence} with cosine similarity calculation (we name it as SentBERT in the following) and \emph{ICRALM}~\cite{ram2023context}, covered in Section~\ref{sec:Evaluation}. \rev{To maintain the integrity of the description of the domain knowledge, we use paragraph-level chunk size in the retrieval process, where each text chunk describes one function or feature (similar to the example in Fig.~\ref{fig:running_example}).} Our approach is flexible for the other similarity metrics in RAG. 

\sectopic{Visual Information Retrieval.}
As discussed in Section~\ref{sec:introduction}, the system's UI details (if available) can be valuable for generating more accurate ACs (as we also evaluate in Section~\ref{subsec:Results}). To retrieve the most closely related visual information (captured as UI screenshots in our case), we use the user story as the input for visual RAG (V-RAG) models. The primary distinction between T- and V-RAG lies in the processing method: instead of textual features, visual information retrieval involves extracting visual features from images to assess their relevance to the query text, which requires a more comprehensive process that incorporates both textual and visual modalities. 

\begin{figure}
    \centering
    \includegraphics[width=0.9\linewidth]{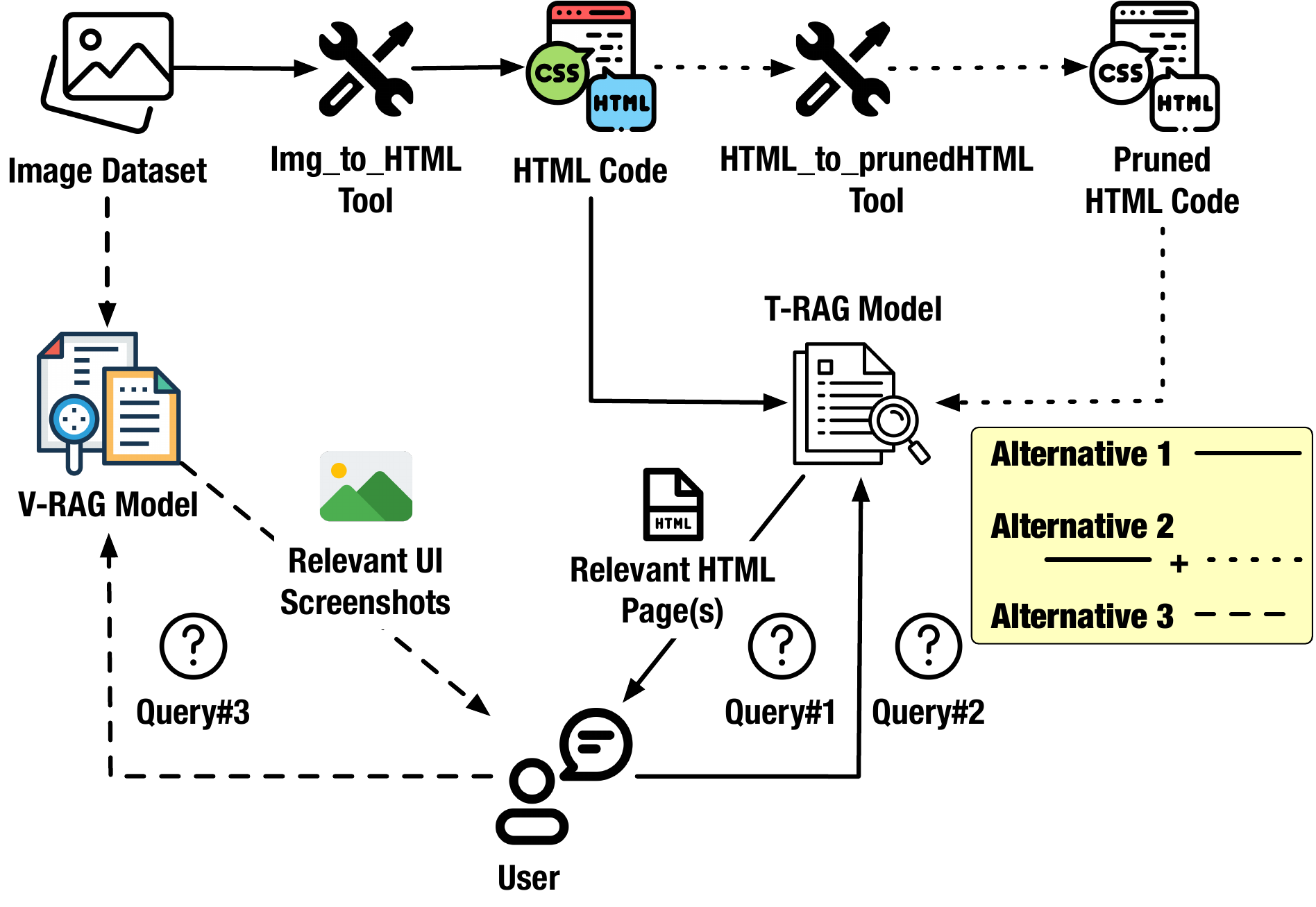}
    \vspace*{-1em}
    \caption{Visual-RAG: Three Alternatives}
    \label{fig:vrag}
    \vspace*{-2em}
\end{figure}

V-RAG offers several possibilities. We have implemented three of these possibilities in our approach (see Fig.~\ref{fig:vrag}). In the first alternative (query\#1), all the visual content is converted into HTML using an image-to-HTML converter, and the best-matching HTML pages are retrieved. The second alternative (query\#2) is an extension of the first alternative, wherein the redundant information in HTML pages is further pruned to retain only the key information, without additional details like CSS. This pruning reduces noise and enhances the specificity and relevance of the retrieved data. In the third alternative (query\#3), visual embeddings from images are directly indexed without HTML conversion. The user's textual query is matched against these visual embeddings, allowing the direct retrieval of visually similar artefacts.

\subsection{Acceptance Criteria Generation}
\begin{figure}
    \centering
    \includegraphics[width=0.9\linewidth]{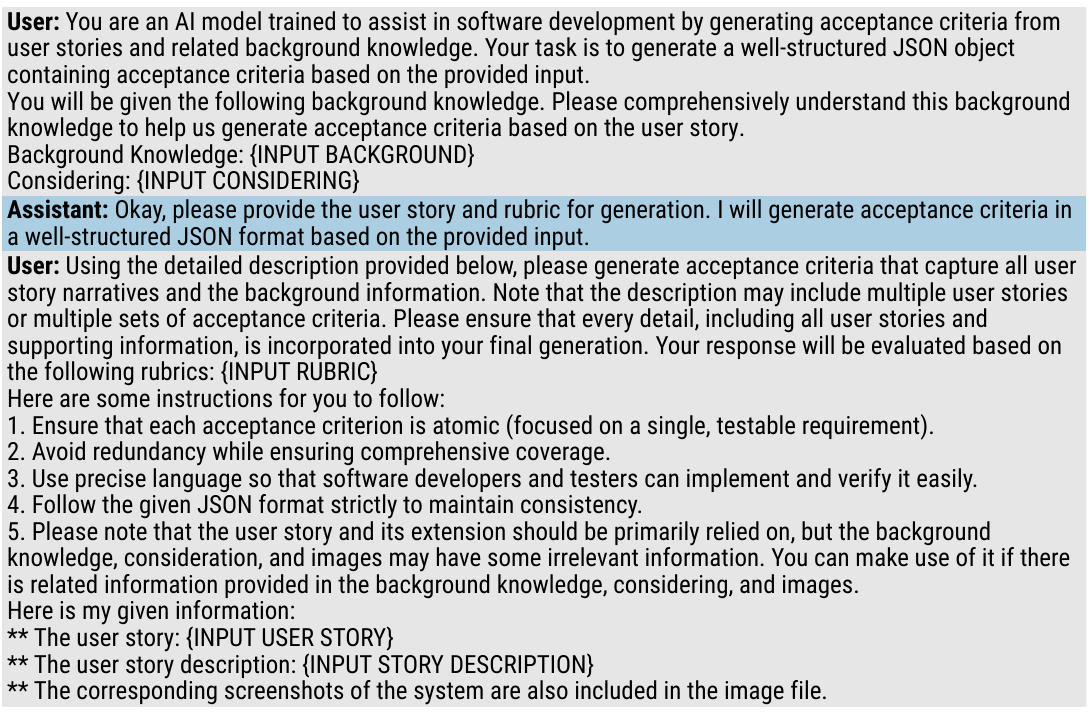}
    \vspace*{-1em}
    \caption{APEER Prompt Template}
    \label{prompt:apeer}
    \vspace*{-2em}
\end{figure}

Once the multi-modal background information is available from the previous step, the next step is to generate ACs originating from the user story. Here, we have two sub-steps, namely prompt construction and generation. Prompt engineering can make a significant difference in the generated outcomes~\cite{arora2024advancing, Vogelsang2025,huang2025prompt}. Chain of Thought (CoT) enables the logical thinking ability of LLMs by providing specific execution steps. We experiment with two different CoT prompt construction methods -- (1) \emph{Urial}~\cite{Liang2023PromptingLL}, which focuses on prompt construction with in-context learning by restyling the given information into structured instruction with in-context alignment, and (2) \emph{APEER}~\cite{jin2024apeer} which focuses on the ground knowledge in prompt construction but without any examples. The example prompt template of APEER is detailed in Fig.~\ref{prompt:apeer}. All other prompt templates are also made available~\cite{ourRepo}. In our framework, we have two types of input, including textual and visual information. As for the multi-modal LLMs, we adopted the image parsing method by directly reading the base64 encoded information of the image.

\subsection{Reward-based Post Processing}
\label{sec:quality_assurance}
While LLMs are employed as generators for AC, their opaque processing process~\cite{cambria2024xai,jiao2024navigating} hinders the guarantee of consistently high-quality outputs. Yet RAG enhances the generation process by providing additional knowledge, introducing noise will also lead to irrelevant or unexpected results, which is named ``retrieval noise'' or ``noise propagation ''~\cite {fang2024enhancing,zhu2024information}. To ensure the quality of the generated AC, we incorporate a post-processing/polishing step using reward models, which refines any low-quality AC and ensures alignment with the given user stories. This thereby improves the overall reliability of the generated AC. The notion of quality in AC (based on best practices and guidance from our industry partner) are:
    
    $\bullet$ Avoid vague AC, like ``The system should work properly.'', as providing minimal guidance to the QA team.
    
    $\bullet$ \rev{AC should be concise and include measurable outcomes for testability, unambiguity, and consistency.}
    
    $\bullet$ Generated AC should be comprehensive, i.e., cover edge cases, and positive and negative outcomes.
    
    $\bullet$ Follow recommended syntax and atomicity, i.e., ``GIVEN, WHEN, THEN'' and provide atomic outcomes for ``Then'' part.


\sectopic{Score-based Polishing.} Algorithm~\ref{alg:refine_ac} outlines the AC polishing process. Our approach leverages reward models that assign scores to the generated candidates based on the user story. Instead of only considering the individual quality of each AC, we introduce an overall quality check for all ACs. We use a global reward model that classifies generated AC into predefined quality levels as the overall assessment. {Prometheus}~\cite{kim2024prometheus} is an open-source reward model that supports absolute rewarding.
\rev{We define five levels (scored 1-5) using a six-dimensional rubric to determine the overall quality of the generated AC. We designed the rubric based on the dimensions in our expert evaluation (see Sec.~\ref {sec:expert}), but with a more atomic description, facilitating the understanding of LLMs.} Given space constraints here, all material is available in our repository~\cite{ourRepo}.

\begin{algorithm}
\caption{Polishing ACs using Reward Models}

\footnotesize
\label{alg:refine_ac}
\begin{algorithmic}[1]

\REQUIRE Set of generated acceptance criteria $\mathcal{AC} = \{ac_1, ac_2, \dots, ac_n\}$, user story $S$, global threshold $T$
\ENSURE Refined set of $\mathcal{AC}'$

\STATE $score_{\text{global}} \leftarrow \text{GlobalRewardModel}(S, \mathcal{AC})$
\IF{$score_{\text{global}} \geq T$}
    \RETURN $\mathcal{AC}$
\ELSE
    \FOR{each $ac_i \in \mathcal{AC}$}
        \STATE $score_i \leftarrow \text{LocalRewardModel}(S, ac_i)$
    \ENDFOR
    \STATE $ac_{\text{worst}} \leftarrow \arg\min_{ac_i \in \mathcal{AC}} score_i$
    \STATE $ac_{\text{polished}} \leftarrow \text{LLM\_Polish}(ac_{\text{worst}}, \mathcal{AC} \setminus \{ac_{\text{worst}}\}, S)$
    \STATE Replace $ac_{\text{worst}}$ in $\mathcal{AC}$ with $ac_{\text{polished}}$
    \RETURN $\mathcal{AC}$
\ENDIF
\end{algorithmic}
\end{algorithm}

We also employ a local reward model to assign a numeric score to each AC. The numeric score is more friendly for us to select the low-quality AC among all the candidates. We experiment with two local reward models, namely {Generative Verifier}~\cite{zhang2024generative} and {UR3}~\cite{yuan-etal-2024-improving}. Given a predefined prompt with the reward objective, generative verifier~\cite{zhang2024generative} will predict the probability of ``yes'' and ``no'' tokens based on this prompt, which is similar to the idea of binary classification. The scale of the probability is from 0-1, where a higher probability of “yes” indicates that the prediction aligns better with the stated objectives. UR3~\cite{yuan-etal-2024-improving}, compared with the generative verifier, prefers a comprehensive estimation of language perplexity for all the input tokens, including the prompt. It quantifies the score estimation process by measuring the divergence using the Kullback–Leibler divergence, which refers to the LLM’s approximation and the actual objective-specific distribution. \textit{Based on the output of the reward model, if ACs do not meet the quality checks, we re-generate the lowest-scored acceptance criterion unless all generated criteria are above a certain level.} \rev{The polishing is not for all generations. We set the threshold for polishing to 5, which means all the ACs with scores less than five will be polished.} Ultimately, we have all the acceptance criteria that meet the quality checks in this post-processing step, as the final output of \approachName.

\section{Empirical Evaluation}
\label{sec:Evaluation}

\subsection{Research Questions}
\label{subsec:RQs}

\sectopic{RQ1. Which \approachName~configuration leads to the most accurate acceptance criteria generation results?} As presented in Section~\ref{sec:approach}, different parts of \approachName~can be instantiated using various alternatives. RQ1 aims to identify the V-RAG and T-RAG alternative techniques and the prompting strategies that yield the most accurate ACs generation results. We further perform an ablation study to assess the advantages (if any) of providing background information for ACs generation. 


\sectopic{RQ2. Which combination of reward models and LLMs yields the best alignment between generated acceptance criteria and the underlying user story?} Reward models are applied to improve the generated AC's structure and alignment with the underlying user story. RQ2 analysis demonstrates how the reward model improves alignment. We also determine the best LLM alternative for generating ACs.


\sectopic{RQ3. Do practitioners find the generated acceptance criteria useful?} In RQ3, we report on an interview survey with practitioners to gain insights into the usefulness of the generated AC from their perspective, based on the best alternatives from RQ1 and RQ2 and previously unseen data.

\subsection{Implementation Details}
\label{subsec:Implementation}
\sectopic{Platform.} We conducted our experiments on an NVIDIA A100 GPU using full-parameter within the PyTorch framework. All models were initialized in bfloat16 and 4-bit mode.

\sectopic{Data Preprocessing.} Our dataset consists of domain documents in paragraph or sentence style, where we restructure domain documents into JSON files.

\sectopic{Information Retrieval.} 
For ICRALM, we use Llama 3B~\cite{Llama3b} with HTML${_\mathrm{Prune}}$ and HTML${_\mathrm{Full}}$. We use an open-source IMG2HTML tool~\cite{Img2html} and a pruning tool in Tan et al.~\cite{tan2024htmlrag}. Additionally, we employ DSE and SentBERT from Hugging Face~\cite{phi3DocmatixV2, allMiniLML12V2}. The default prompts for these methods are used, based on their respective repositories.

\sectopic{Prompts and LLMs Configuration.}
We adopt Urial and APEER prompt templates, as detailed in their appendices. For image parsing, we directly process images using base64 encoding. We utilize the latest versions of each model available before March 2025, with default hyperparameter settings, including temperature and top\_p, for initialization, but maintain the maximum input length to ensure flexibility in processing.

\sectopic{Reward Models and Polishing.}
Prometheus~\cite{kim2024prometheus} provides fine-tuned checkpoints on Hugging Face. The Generative Verifier is adaptable to any open-source LLM. We initialize them using Llama 3.1 8B~\cite{Llama8bInstruct}. For polishing, we maintain the same LLM settings as in the previous step. The polishing prompt is appended to the AC generation prompt, creating a multi-cycle dialogue to maintain contextual understanding.

\subsection{Data Collection Procedure.}
\label{subsec:DataCollection} 
The data collection procedure in our industry study consists of two stages: collecting industry data and gathering expert feedback. In the first stage, we gathered user stories, ACs, and the relevant multi-modal documents. In the second stage, we return to experts from \company~to obtain their evaluation of ACs generated by \approachName.

\sectopic{Industry Data Collection.} We collected the data from Jira~\cite{Jira} using official REST API, where each case of the data consists of \emph{User Story and Extension}, \emph{Background Description}, \emph{Consideration} \rev{(system requirements or non-functional requirements)}, \emph{Screenshots}, and \emph{Acceptance Criteria}. The user story contains the necessary information as input. The background description, consideration, and screenshots are the background knowledge, which are structured into textual and visual databases for querying (see Figure~\ref{fig:framework}). Table~\ref{tab:case_des} provides a brief description of the two samples, referred to as Sample~A and Sample~B. Sample~A has been used in our approach to select different RAG, LLM, and reward model alternatives in RQs 1 and 2. Sample~B data has been used to evaluate the best combination of \approachName~for expert evaluation in RQ3. We intentionally collected expert feedback on previously unseen data (Sample B) to prevent evaluation bias, since the determined combination on Sample A should not be used with the same data for expert evaluation. Therefore, evaluating on separate data ensures an objective assessment of the effectiveness.

\begin{table}[htbp]
    \vspace{-1em}
    \centering
    \caption{Dataset Description}
    \vspace{-1em}
    \label{tab:case_des}
    \setlength{\tabcolsep}{1pt}
    \resizebox{\linewidth}{!}{
    \begin{tabular}{@{}|c|c|c|c|l|@{}}
        \hline
        \rowcolor[HTML]{E0E0E0} 
        \textbf{\begin{tabular}[c]{@{}c@{}}\small Sample\end{tabular}} & 
        \textbf{\begin{tabular}[c]{@{}c@{}}\small User Stories\end{tabular}} & 
        \textbf{\begin{tabular}[c]{@{}c@{}}\small ACs\end{tabular}} & 
        \textbf{\begin{tabular}[c]{@{}c@{}}\small Atomic ACs\end{tabular}} & 
        \multicolumn{1}{c|}{\textbf{\small Description}} \\
        \hline
        A & 42 & 169 & 224 & \projName~course structuring for learners. \\
        \hline
        B & 32 & 109 & 179 & \projName~overall system management for all users. \\
        \hline
    \end{tabular}
    }
    \vspace{-1em}
\end{table}


\sectopic{Expert Feedback Collection.}~\label{sec:expert}
In this stage, we collected data from three experts from \company, in an interview survey to assess their perception of the quality of the generated ACs in RQ3. The three experts include a senior business analyst (Expert 1), a functional integration business analyst (Expert 2), and a senior quality assurance and solutions architect (Expert 3). Expert 1 has three years of experience in \projName~and 10 years of total industry experience, Expert 2 has 2.5 years of \projName~experience and 7 years in the software industry, and Expert 3 has worked at \company~for 4 years and 15 years in total in the SE industry. \rev{We only recruit the experts from \company~rather than widely finding the experts from open domains due to their professional knowledge and understanding about \projName.}

The interview survey session lasted $\approx$2 hours. At the beginning of the session, we explained the evaluation procedure to the experts and provided a quick recollection of the project requirements. Given the limited availability of experts, we were able to cover detailed feedback on 17 user stories and 81 corresponding ACs generated using \approachName~for Sample~B, in a random order. The experts reviewed each AC at a time and had access to the corresponding user story. For each AC and a user story, we asked experts to rate them on three quality aspects (noted below). For all ACs generated for a given user story, they also rated their \emph{coverage}. The experts rated each of these four aspects on a five-point Likert scale~\cite{likert1932technique}. 
Rated 1 (Strongly Disagree) means the AC significantly fails to meet the quality aspect. Rated 2 ({Disagree}) means that the AC does not adequately meet the quality aspect.
Rated 3 ({Neutral}) means that the AC somewhat meets the quality aspect, but improvements are necessary.
Rated 4 ({Agree}) means that the AC meets the quality aspect satisfactorily, with only minor issues or omissions.
Rated 5 ({Strongly Agree}) means that the AC excellently meets the quality aspect. The four quality aspects, relevance, correctness and understandability for each AC and coverage for all ACs of a given user story are described below:

    $\bullet$ \emph{Relevance} assesses how well the generated AC aligns with the given user story.
    
    $\bullet$ \emph{Correctness} evaluates the accuracy of AC, format and the information captured in the AC. In short, the AC should be factually accurate and logically structured.
    
    $\bullet$ \emph{Understandability} assesses whether the generated AC is clear and comprehensible to the experts without any redundancies or inconsistent terminology introduced by an LLM.
    
    $\bullet$ \emph{Coverage} evaluates the ‘completeness’ of the generated ACs, i.e., which ACs encompass all relevant aspects of the user story, e.g., positive \& negative flows and edge cases.

We asked experts to share their rationale for a given rating on each AC. The experts were free to agree or disagree with, or even change their rating, in case another expert had a different rationale that they might have missed. Ultimately, a consensus rating was achieved for each AC. We noted down the ACs where they did not agree on a rating. We also noted down each expert's brief sharing from their overall perspective on the results at the end of the session.


\subsection{Evaluation Procedure and Metrics}
\label{subsec:EvalProcedure}
There are eight main independent considerations in our approach, namely, (i) T-RAG model, (ii) $k$ value for T-RAG (i.e., the number of textual passages retrieved by T-RAG model), (iii) V-RAG model, (iv) $k$ value for V-RAG (i.e., the number of visual documents retrieved by V-RAG model), (v) prompt construction techniques, (vi) selected LLMs, (vii) local reward models (note that we maintain global reward model as an integral part of our approach); and (viii) choice of using background information or not, or only using textual background. Table~\ref{tab:choices} shows all the choices for these. While an exhaustive evaluation of all possible combinations would be ideal, practical constraints necessitate a selective approach. Therefore, the alternative models chosen for each case are motivated either by their widespread adoption in NLP literature~\cite{zhao2024retrieval,gao2023retrieval,li2022survey} or by the availability of implementations. \rev{Given the ground truth from \projName~data, we conducted a stepwise evaluation in RQ1 and RQ2 to determine the best configuration for \approachName in three steps, including 1) information retrieval, 2) AC generation, and 3) AC polishing. while the generation ability is strongly determined by the generative methods, such as LLMs, Pretrained LMs, etc, they separately evaluate the information retrieval process for the RAG procedure~\cite{fang2019deep,xie2023survey,ou2022counterfactual}. We adopt a similar strategy to separately evaluate the information retrieval results and the AC generation results.}


\begin{table}[]
    \vspace{-1em}
    \centering
    \caption{Choices Table}
    \vspace{-1em}
    \label{tab:choices}
    \setlength{\tabcolsep}{1.5pt}
    \footnotesize 
    \renewcommand{\arraystretch}{1} 
    \resizebox{0.9\linewidth}{!}{
    \begin{tabular}{@{}|>{\centering\arraybackslash}m{2.4cm}|>{\centering\arraybackslash}m{0.85cm}|>{\raggedright\arraybackslash}p{5.8cm}|@{}}
    \hline
        \rowcolor[HTML]{E0E0E0} 
        \textbf{\begin{tabular}[c]{@{}c@{}}Independent\\Variable\end{tabular}} & 
        \textbf{\begin{tabular}[c]{@{}c@{}}Num\_\\Choices\end{tabular}} & 
        \textbf{Choices}\\ \hline
        T-RAG & 2 & \begin{tabular}[c]{@{}c@{}}\emph{SentBERT}~\cite{reimers2019sentence} and \emph{ICRALM}~\cite{ram2023context}\end{tabular} \\ \hline
        T-RAG ($k$) & 4 & \begin{tabular}[c]{@{}c@{}}1, 5, 10 and 20\end{tabular} \\ \hline
        V-RAG & 3 & \begin{tabular}[c]{@{}c@{}} \emph{HTML$_\mathrm{Prune}$}, \emph{HTML$_\mathrm{Full}$} and \\\textit{Document Screenshot Embedding (DSE)}~\cite{phi3DocmatixV2}\end{tabular} \\ \hline
        V-RAG ($k$) & 4 & \begin{tabular}[c]{@{}c@{}}1, 5, 10 and 20\end{tabular} \\ \hline
        Prompt Construction & 2 & \begin{tabular}[c]{@{}c@{}}\emph{Urial}~\cite{Liang2023PromptingLL} and \emph{APEER}~\cite{jin2024apeer}\end{tabular} \\ \hline
        LLMs & 3 & \begin{tabular}[c]{@{}c@{}}\emph{Claude}~\cite{Claude}, \emph{Gemini}~\cite{Gemini} and \emph{GPT4o}~\cite{GPT4o}\end{tabular} \\ \hline
        Local Reward Model & 2 & \begin{tabular}[c]{@{}c@{}}\emph{Generative Verifier}~\cite{zhang2024generative} and \emph{UR3}~\cite{yuan-etal-2024-improving}\end{tabular} \\ \hline
        Background Information & 3 & \begin{tabular}[c]{@{}c@{}}With VRAG and TRAG, \\No VRAG or TRAG (w/o RAG), and \\No VRAG only TRAG (w/o VRAG)\end{tabular} \\ \hline
    \end{tabular}
    }
    \vspace{-2em}
\end{table}



\subsubsection{Evaluation Metrics - Information Retrieval}
The general results of information retrieval will return a list of the ranked candidates. Following the previous RAG and recommendation work in the NLP domain, which all related to the ranking task, we adopted similar evaluation metrics for the ranking results as \textbf{Precision@K}, \textbf{Recall@K}, \textbf{F1-Score@K}, \textbf{NDCG@K}, \textbf{HitRate@K} and \textbf{MAP}~\cite{fang2019deep,xie2023survey}. These metrics are commonly used in evaluating ranking and recommendation systems. Precision@K measures the proportion of the relevant items in the top-$k$ recommended items, focusing on the accuracy of the top suggestions. Recall@K assesses the fraction of all relevant items that appear in the top k recommendations, indicating how well the system retrieves relevant items overall. F1@K is the harmonic mean of precision@k and recall@k, providing a single metric that balances both precision and recall in the top-$k$ results. MAP (Mean Average Precision) averages the precision scores obtained at the ranks where relevant items occur across multiple queries, reflecting the overall ranking quality. NDCG@K (Normalized Discounted Cumulative Gain) evaluates the ranking quality by considering the position of relevant items, giving higher scores when relevant items are ranked higher, and normalizing the score to allow comparisons. Finally, HitRate@K measures the percentage of users or queries for which at least one relevant item appears in the top-$k$ results, offering a simple indicator of whether the system makes any correct recommendations at all.

\begin{algorithm}
\footnotesize
\caption{Evaluation of ACs Using LLMs as Judges}
\begin{algorithmic}[1]
\label{alg:llm_evaluation}
\REQUIRE Generated AC, ground truth ACs split into test objectives $\mathcal{T} = \{T_1, T_2, \dots, T_n\}$, user stories, and LLM judges $\mathcal{J} = \{\text{Claude 3.5 Sonnet},\ \text{Gemini 2 Flash},\ \text{GPT4o}\}$

\ENSURE Accuracy metrics: Acc@Hit and Acc@Cor.

\STATE Initialize hit counter $hit \leftarrow 0$ and correct counter $correct \leftarrow 0$
\FOR{each test objective $T_i \in \mathcal{T}$}
    \STATE Retrieve the corresponding generated ACs segment for $T_i$
    \FOR{each LLM judge $J \in \mathcal{J}$}
        \STATE Query $J$ with the ACs segment and $T_i$
        \STATE Receive judgment $j \in \{\text{Full Cov.},\ \text{Partial Cov.},\ \text{Not Cov.}\}$
    \ENDFOR
    \IF{all judges return at least \textit{Partial Coverage}}
        \STATE $hit \leftarrow hit + 1$ \COMMENT{Acc@Hit: test objectives are partially covered}
    \ENDIF
    \IF{all judges return \textit{Full Coverage}}
        \STATE $cor. \leftarrow cor. + 1$ \COMMENT{Acc@Cor.: test objectives are fully covered}
    \ENDIF
\ENDFOR
\STATE Compute accuracy metrics:
\STATE $\text{Acc@Hit} \leftarrow \frac{hit}{n}$
\STATE $\text{Acc@Cor.} \leftarrow \frac{correct}{n}$
\RETURN $\text{Acc@Hit},\ \text{Acc@Cor.}$

\end{algorithmic}
\end{algorithm}

\subsubsection{Evaluation Metrics - AC Generation}
\rev{We adopt two types of evaluation metrics for AC generation. \textbf{Statistical Evaluation Metrics.} Inspired by previous work~\cite{arora2024generating}, we adopted similar statistical measures for acceptance criteria generation. These include \textit{Semantic Similarity} using SentBERT~\cite{reimers2019sentence}, \textit{ROUGE-1}, \textit{ROUGE-2}, \textit{ROUGE-L}, \textit{BLEU}, and \textit{Levenshtein} \textit{distance}~\cite{borges2019combining,hu2024unveiling,hanke2024open}. \textit{Semantic Similarity} embeds each sentence and scores them with cosine similarity \cite{reimers2019sentence}. \textit{ROUGE-1} / \textit{ROUGE-2} / \textit{ROUGE-L} match single words, word pairs, and the longest shared word sequence, respectively. \textit{BLEU} counts n-gram matches and adds a brevity penalty to discourage very short outputs. \textit{Levenshtein distance} counts the fewest single-character edits needed to turn one string into the other. \textbf{LLMs-as--judges from SE Perspective.}~\label{sec:llms_as_judges}LLM-as-a-judge is commonly adopted in the evaluation of generative tasks, including RAG~\cite{es2024ragas,wei2024systematic}. However, they mostly execute on the semantic level and lack adaptation in specific domains. We introduce the LLMs-as-judges metric by manually segmenting the ACs to atomic objectives to ensure the AC-level evaluation (original single AC will cover several objectives, which will yield inconsistent evaluation results) and clearly defining the instructions written by software testing experts (e.g., evaluation rubric, AC definitions, etc). We manually segment the ACs into distinct points for each test objective in the ground truth, i.e., in Fig.~\ref{fig:running_example}'s example, there are three objectives in the second AC, including ``(1) I should see..'', ``The widget should have...'', and ``The widget should show...'', we manually segment this AC into three ACs. The LLMs then assess whether these objectives are covered in the generated criteria on a point-by-point basis. We define three-level coverage for the atomic testing objectives, where Full Cov. refers to all the segmented testing objectives that are covered in the generated ACs, Partial Cov. refers that the test objectives are mentioned but not fully covered, Not Cov. means the test objectives are completely not covered. Our evaluation framework defines two accuracy metrics: \textbf{Acc@Hit}, which measures partial coverage of test objectives, and \textbf{Acc@Cor.}, which indicates full coverage. Since ACs are not isolated from user stories, we further introduce two evaluation types to analyze the results from different perspectives. \textbf{Case acc. (C)} is the average accuracy calculated per user story, reflecting the model’s generalization ability across different user stories. A higher case accuracy suggests stronger adaptability to varying contexts. \textbf{Point acc. (P)} measures accuracy for individual test objectives within the ground truth ACs, representing the absolute performance of the generation process. Variations between case and point accuracy reveal different tendencies in method performance. For instance, if a method achieves high point accuracy but relatively low case accuracy, it suggests that while the method can generate precise criteria for some user stories, it fails to generalize effectively across others. To mitigate potential bias in evaluation, we employ three LLMs and set temperature and top\_p as 0 and 0.1 to maximally reduce the randomness. The details are specified in Algorithm~\ref{alg:llm_evaluation}, where the positive judgment is recorded only when all models \emph{unanimously} agree.}

\subsubsection{Evaluation Metrics - AC Polishing}
We primarily adopt the evaluation metrics outlined in the AC generation process. Since polishing is not applied to all user stories, we introduce an additional LLM-annotated evaluation to compare the original and polished ACs. In this evaluation, we present the user story, original ACs, and polished ACs to three LLMs and ask them to determine which version is superior based on a predefined rubric. Compare accuracy is measured by only counting where all three LLMs \emph{unanimously} judge the polished version is comparable better, similarly to the judgment methodology described in Section~\ref{sec:llms_as_judges}.

\begin{table*}[]
\vspace{-1em}
\centering
\caption{Results of Information Retrieval (RQ1)}
\vspace{-1em}
\label{tab:rq1_rag}
\resizebox{\textwidth}{!}{%
\begin{tabular}{l|c|ccccc|ccccc|ccccc|ccccc}
\hline
\multicolumn{1}{c|}{\multirow{2}{*}{\textbf{Methods}}} & \multirow{2}{*}{\textbf{MAP}} & \multicolumn{5}{c|}{\textbf{@K=1}}                                  & \multicolumn{5}{c|}{\textbf{@K=5}}                                  & \multicolumn{5}{c|}{\textbf{@K=10}}                                 & \multicolumn{5}{c}{\textbf{@K=20}}                                  \\ \cline{3-22} 
\multicolumn{1}{c|}{}                                  &                               & \textbf{P} & \textbf{R} & \textbf{F1} & \textbf{nDCG} & \textbf{HR} & \textbf{P} & \textbf{R} & \textbf{F1} & \textbf{nDCG} & \textbf{HR} & \textbf{P} & \textbf{R} & \textbf{F1} & \textbf{nDCG} & \textbf{HR} & \textbf{P} & \textbf{R} & \textbf{F1} & \textbf{nDCG} & \textbf{HR} \\ \hline
SentBERT                                               & 10.35                         & 5.56       & 5.56       & 5.56        & 5.56          & 5.56        & 2.22       & 11.11      & 3.70        & 9.06          & 11.11       & 2.22       & 22.22      & 4.04        & 12.47         & 22.22       & 1.67       & 33.33      & 3.17        & 15.21         & 33.33       \\
ICRALM                                                 & \textbf{39.65}                         & \textbf{33.33}      & \textbf{33.33}      & \textbf{33.33}       & \textbf{33.33}         & \textbf{33.33}       & \textbf{8.33}       & \textbf{41.67}      & \textbf{13.89}       & \textbf{38.23}         & \textbf{41.67}       & \textbf{5.56}       & \textbf{55.56}      & \textbf{10.10}       & \textbf{42.63}         & \textbf{55.56}       & \textbf{3.47}       & \textbf{69.44}      & \textbf{6.61}        & \textbf{46.00}         & \textbf{69.44}       \\ \hline
DSE                                                   & \textbf{15.71} & \textbf{11.11} & \textbf{5.56} & \textbf{6.94} & \textbf{11.11} & \textbf{11.11} & \textbf{7.22} & 19.21 & 9.88 & 15.04 & \textbf{27.78} & \textbf{6.11} & 29.86 & \textbf{9.77} & \textbf{19.29} & 36.11 & \textbf{4.58} & \textbf{}38.89 & \textbf{8.03} & \textbf{22.56} & 38.89 \\
HTML\(_{\mathrm{Prune}}\)                              & 11.97 & 5.56 & 3.70 & 4.17 & 5.56 & 5.56 & 4.44 & 15.28 & 6.58 & 10.56 & 19.44 & 4.17 & 25.69 & 6.92 & 14.54 & 30.56 & 3.89 & \textbf{41.67} & 6.95 & 19.62 & \textbf{41.67} \\
HTML\(_{\mathrm{Full}}\)                               &  12.82 & 5.56 & 2.31 & 3.24 & 5.56 & 5.56 & \textbf{7.22} & \textbf{19.68} & \textbf{9.92} & 13.04 & \textbf{27.78} & 5.56 & \textbf{30.56} & 9.03 & 17.00 & \textbf{41.67} & 4.44 & 41.67 & 7.85 & 20.95 & \textbf{41.67} \\ \hline
\end{tabular}%
}
\vspace{-1em}
\end{table*}
\begin{table*}[]
\footnotesize
\centering
\caption{Results of Acceptance Generation Using LLMs (RQ1)}
\label{tab:rq1_ac}
\vspace*{-1em}
\resizebox{0.75\linewidth}{!}{
\begin{tabular}{l|cccc|cccccc}
\hline
\multicolumn{1}{c|}{\multirow{2}{*}{\textbf{Methods}}} & \multicolumn{4}{c|}{\textbf{LLMs-Annotated}}                                          & \multicolumn{6}{c}{\textbf{Statistical}}                                                                                                     \\ \cline{2-11} 
\multicolumn{1}{c|}{}                                  & \textbf{Hit(C)}         & \textbf{Cor(C)} & \textbf{Hit(P)}         & \textbf{Cor(P)} & \textbf{Sim.}           & \textbf{Leven.} & \textbf{BLEU}   & \textbf{ROUGE-1}         & \textbf{ROUGE-2}         & \textbf{ROUGE-L}         \\ \hline
Claude~$_{\mathrm{Urial}}$                             & \textit{61.81}          & \textit{13.67}  & \textit{57.02}          & \textit{12.40}  & \textit{\textbf{72.86}} & \textit{1525}   & 0.1985          & \textit{0.3637}          & \textit{0.1448}          & \textit{0.2956}          \\
Claude~$_{\mathrm{APEER}}$                             & 61.80                   & 14.91           & 56.20                   & 14.88           & 71.23                   & 1520            & \textbf{0.2271} & 0.3651                   & 0.1433                   & 0.2832                   \\
Claude~$_{\mathrm{Urial~w/o~VRAG}}$                          & 57.48                   & 12.97           & 53.93                   & 10.20           & 71.94                   & \textit{1525}   & \textit{0.2042} & 0.3573                   & 0.1417                   & 0.2927                   \\
Claude~$_{\mathrm{w/o~RAG}}$                           & 48.72                   & 5.62            & 37.60                   & 3.72            & 70.64                   & 1576            & 0.1830          & 0.3009                   & 0.1004                   & 0.2333                   \\ \hline
Gemini~$_{\mathrm{Urial}}$                             & \textit{\textbf{71.19}} & \textit{15.95}  & \textit{\textbf{68.60}} & \textit{19.01}  & \textit{72.18}          & 1503            & 0.2161          & \textit{\textbf{0.3940}} & \textit{\textbf{0.1571}} & \textit{\textbf{0.3682}} \\
Gemini~$_{\mathrm{APEER}}$                             & 69.93                   & \textbf{29.19}  & 64.05                   & \textbf{24.38}  & 71.24                   & 1575            & 0.2209          & 0.3591                   & 0.1401                   & 0.3169                   \\
Gemini~$_{\mathrm{Urial~w/o~VRAG}}$                          & 64.76                   & 13.69           & 57.85                   & 14.53           & 71.97                   & \textit{1496}   & \textit{0.2206} & 0.3883                   & 0.1560                   & 0.3626                   \\
Gemini~$_{\mathrm{w/o~RAG}}$                           & 46.37                   & 10.32           & 34.71                   & 7.44            & 71.03                   & 1742            & 0.1653          & 0.3102                   & 0.1041                   & 0.2818                   \\ \hline
GPT4o~$_{\mathrm{Urial}}$                              & 62.22                   & \textit{13.31}  & 56.20                   & \textit{13.64}  & 71.75                   & \textbf{1491}   & 0.1744          & 0.3796                   & 0.1479                   & 0.3540                   \\
GPT4o~$_{\mathrm{APEER}}$                              & 60.79                   & 12.11           & 56.20                   & 15.70           & 71.02                   & 1505            & 0.2037          & 0.3355                   & 0.1158                   & 0.2803                   \\
Gemini~$_{\mathrm{Urial~w/o~VRAG}}$                          & \textit{63.31}          & 9.93            & \textit{57.85}          & 8.64            & \textit{71.93}          & 1498            & \textit{0.2109} & \textit{0.3847}          & \textit{0.1561}          & \textit{0.3561}          \\
GPT4o~$_{\mathrm{w/o~RAG}}$                            & 48.06                   & 8.31            & 35.54                   & 5.79            & 71.16                   & 1550            & 0.1644          & 0.2954                   & 0.0795                   & 0.2464                   \\ \hline
\end{tabular}%
}
\vspace*{-2em}
\end{table*}
\subsection{Results and Discussion}
\label{subsec:Results}

\sectopic{RQ1.}~\label{subsec:RQ1}
Table~\ref{tab:rq1_rag} presents the results of experiments with T-RAG and V-RAG alternatives (Table~\ref{tab:choices}). For T-RAG alternatives, ICRALM consistently outperforms SentBERT across all $k$ values, i.e., in terms of retrieving the relevant textual documents, ICRALM performs the best. As for V-RAG models, three alternatives achieve similar results, where DSE is slightly better than HTML$_\mathrm{Full}$, followed by HTML$_\mathrm{Prune}$. During our retrieval process, HTML$_\mathrm{Full}$, due to the length of the HTML code, has a longer processing time compared with the other two methods. Thus, we select DSE as the best alternative for V-RAG. Besides, we need to determine the number $k$ in top-$k$ settings, while we cannot pass all the top information to the generation step. Considering the input size for LLMs, we set top-$5$ for the textual and visual information retrieval. For ICRALM,  top-$5$ results enable a balanced trade-off, which allows ICRALM to have acceptable recall (41.67) compared with the conditions of $@k=10$ and $@k=20$ while maintaining acceptable precision (8.33) compared with the condition of $@k=1$. For V-RAG, we observed that with the increase in $@k$, the precision does not decrease significantly, but the recall incrementally increases. DSE approaches yield their highest F1 scores at  $@k=5$, indicating the best compromise between retrieving enough relevant documents (boosted recall) and avoiding the dilution of precision that occurs with higher $@k$ values. Thus, \textit{we use ICRALM for T-RAG with top-$5$ results and DSE for V-RAG with top-$5$ results.}

Table~\ref{tab:rq1_ac} presents the results of ACs generation using ICRALM and DSE, where italicized values indicate improvements on the ablation settings. We tested two prompt templates across three LLMs and included ablation results. In the w/o RAG setting, we retained the Urial prompt template but removed the image input. The findings in Table~\ref{tab:rq1_ac} indicate several key observations: (1) The statistical results across all baselines are very similar, suggesting that statistical evaluations alone are insufficient to distinguish performance differences among the baselines; (2) Among the three LLM configurations, the versions incorporating RAG significantly outperform those without RAG, highlighting the importance of RAG integration in the ACs generation process; (3) Comparing the Urial w/o RAG version with the Urial version, we observe that Urial consistently outperforms Urial w/o RAG across most evaluation measures, particularly in correct accuracy. A potential explanation for this is that the textual background description provides sufficient contextual information to guide the generation process, but the inclusion of image input plays a crucial role in ensuring greater accuracy in ACs generation; (4) Of the three LLMs tested, Gemini 2 Flash achieves the best results across all prompt settings, including Urial, APEER, and w/o RAG. GPT-4o and Claude 3.5 Sonnet demonstrate comparable performance, following Gemini 2 Flash; (5) Between the two prompt templates, Urial performs better in terms of Hit@Accuracy, while APEER excels in Correct@Accuracy. This suggests that Urial leads to better relevance in generated ACs, whereas APEER enhances correctness; (6) Case-level accuracy is generally higher than point-level accuracy. Since different user stories correspond to varying numbers of ACs, this suggests that LLMs perform better on smaller cases rather than on complete, more complex ones; (7) Within each LLM, most statistical metrics, such as ROUGE-1, -2, and -L, and BLEU, exhibit trends similar to those observed in LLM-annotated evaluations.


\vspace{-0.5em}

\find{\textbf{Answer to RQ1:} {With k$=$5, ICRALM (T-RAG) and DSE (V-RAG) give the best retrieval: $+$25\% textual F1, $+$15\% visual F1, and $\sim$50\% less runtime than $HTML_{Full}$. Passing these top-5 hits to the generators, RAG boosts case-level accuracy 42.8$\to$66.0\% (+23.2 pp) and point-level accuracy 36.4 → 57.9\% (+21.5 pp). Gemini 2 Flash tops the LLMs; Urial is most relevant (Hit@Acc 74.1\%), while APEER is most correct (Correct@Acc 61.3\%). Overall, the ICRALM$+$DSE pipeline lifts AC-generation F1 by 27\% and accuracy by $>$20 pp vs the strongest non-RAG baseline.}
}
\vspace{-0.5em}


\begin{table*}[]
\vspace{-1em}
\centering
\caption{Results of Acceptance Criteria Polishing (RQ2)}
\label{tab:rq2_polish}
\vspace*{-1em}
\resizebox{0.8\linewidth}{!}{
\begin{tabular}{l|ccccc|cccccc}
\hline
\multicolumn{1}{c|}{\multirow{2}{*}{\textbf{Methods}}} & \multicolumn{5}{c|}{\textbf{LLMs-Annotated}}                                             & \multicolumn{6}{c}{\textbf{Statistical}}                                                                    \\ \cline{2-12} 
\multicolumn{1}{c|}{}                                  & \textbf{Hit(C)} & \textbf{Cor(C)} & \textbf{Hit(P)} & \textbf{Cor(P)} & \textbf{Compare} & \textbf{Sim.}  & \textbf{Leven.} & \textbf{BLEU}   & \textbf{ROUGE-1} & \textbf{ROUGE-2} & \textbf{ROUGE-L} \\ \hline
Claude~$_{\mathrm{Urial}}$                             & 61.81           & 13.67           & 57.02           & 12.40           & NA               & 72.86          & 1525            & 0.1985          & 0.3637           & 0.1448           & 0.2956           \\
Claude~$_{\mathrm{Urial~w/~UR3}}$                             & 69.08           & 16.68           & 62.05           & 14.73           & \textbf{100.00}  & 72.28          & 1540            & 0.1993 & 0.3597           & 0.1430           & 0.2907           \\
Claude~$_{\mathrm{Urial~w/~Gen}}$                             & 68.68           & 15.19           & 61.61           & 13.39           & 83.33            & \textbf{72.95} & 1525            & 0.2023          & 0.3628           & 0.1456           & 0.2946           \\ \hline
Gemini~$_{\mathrm{Urial}}$                             & 71.19           & 15.95           & 68.60           & 19.01           & NA               & 72.18          & 1503            & 0.2161          & \textbf{0.3940}  & \textbf{0.1571}  & \textbf{0.3682}  \\
Gemini~$_{\mathrm{Urial~w/~UR3}}$                             & \textbf{79.10}  & \textbf{17.72}  & \textbf{74.11}  & \textbf{20.54}  & \textbf{100.00}  & 71.79          & 1517            & 0.2180          & 0.3892           & 0.1529           & 0.3660           \\
Gemini~$_{\mathrm{Urial~w/~Gen}}$                             & \textbf{79.10}  & \textbf{17.72}  & \textbf{74.11}  & \textbf{20.54}  & \textbf{100.00}  & 71.80          & 1511            & \textbf{0.2181} & 0.3865           & 0.1545           & 0.3612           \\ \hline
GPT4o~$_{\mathrm{Urial}}$                              & 62.22           & 13.31           & 56.20           & 13.64           & NA               & 71.75          & 1491            & 0.1744          & 0.3796           & 0.1479           & 0.3540           \\
GPT4o~$_{\mathrm{Urial~w/~UR3}}$                              & 69.13           & 15.19            & 60.71           & 15.18           & 88.89            & 70.97          & \textbf{1483}   & 0.1775          & 0.3807           & 0.1455           & 0.3501           \\
GPT4o~$_{\mathrm{Urial~w/~Gen}}$                              & 71.91           & 14.79           & 64.73           & 14.73           & \textbf{100.00}  & 71.39          & 1501            & 0.1749          & 0.3699           & 0.1396           & 0.3483           \\ \hline
\end{tabular}%
}
\vspace*{-2em}
\end{table*}

\sectopic{RQ2.}~\label{sec:rq2}Table~\ref{tab:rq2_polish} presents the polishing results of three LLMs using the Urial template. Our method incorporates two alternative reward models, UR3 and the Generative Verifier. We continue to experiment with all three LLMs in RQ2. Since the global reward model determines whether polishing is necessary, only a subset of user stories ($\approx$20\%) undergoes refinement, representing a small portion of the overall dataset. The results indicate that both UR3 and the Generative Verifier consistently improve the ACs in LLM-annotated evaluations. However, statistical results show inconsistent improvements. Specifically, (1) at the case level, improvements in Hit accuracy are comparable between UR3 and the Generative Verifier across all three LLMs, but UR3 demonstrates relatively greater enhancement in Correct accuracy; (2) At the point level, a similar trend is observed: UR3 yields higher Correct accuracy, while the Generative Verifier enhances the overall relatedness of the generated ACs; (3) Pairwise comparisons between the original and polished versions suggest that, from the LLMs’ perspective, polishing generally improves the overall quality of the ACs; (4) Statistical results remain inconsistent, mirroring the trend observed in Table~\ref{tab:rq1_ac}.

\vspace{-0.5em}
\find{\textbf{Answers to RQ2:} Applying polishing to the stories raises case-level accuracy by $\sim$7 pp and point-level accuracy by $\sim$5 pp. UR3 is the best choice when correctness is paramount, while the Generative Verifier is slightly better for relevance.  With either reward model, Gemini 2 Flash remains the top LLM. Statistical metrics change little, reinforcing that human- or LLM-based judgments are needed to capture the true gains.}
\vspace{-0.5em}

\sectopic{RQ3.} Expert assessment of 81 ACs across 17 user stories measured four dimensions: relevance, correctness, understandability (per AC), and coverage (per user story). Below shows response frequencies on a 5-point scale.
\\
$\bullet$ \textbf{Relevance} [0, 3, 21, 31, 26] (average = 3.99) \\
$\bullet$ \textbf{Correctness} [3, 7, 14, 25, 32] (average = 3.94) \\
$\bullet$ \textbf{Understandability} [0, 0, 20, 24, 37]  (average = 4.21) \\
$\bullet$ \textbf{Coverage} [0, 3, 4, 6, 4] (average = 3.89)
\\
Experts rated all three criteria near 4 (``Agreed''), indicating that \approachName's ACs are generally relevant, correct, and understandable. (1) This shows that the experts agree to a large extent that, on average, the AC generated by our approach is relevant to the underlying user story, correct, and understandable. (2) The correctness score is slightly low, which was due to redundancy in some generated ACs, i.e., LLMs will generate redundant ACs, essentially paraphrased versions of each other. This affects the correctness rating of all such ACs. In the three cases at correctness level 1, terminological issues were found in the generated criteria, and two were redundant in relation to a given user story. (3) Other issues related to ACs being low-rated on relevance, correctness, and understandability included not being particular and capturing the ACs at a very generic level. (4) In terms of coverage, in 7/17 user stories, the experts felt that the generated ACs failed to cover all cases that they would have wanted. In most cases, it was related to edge cases and negative inputs. 

Qualitatively, experts were often impressed: Expert 1 noted ``it's amazing how your system is able to capture bento box,'' a concept of UI container absent from the user story but surfaced by \approachName with RAG; Expert 2 said some ACs clarified the domain and ``It's impressive and would be really useful for a new BA (business analyst).'' They also expected time and effort savings. Expert 3 remarked, ``This also gives some ACs which even as an expert I wouldn't have thought of.'' All expressed interest in deploying the approach, underscoring strong enthusiasm and practical utility.

\find{\textbf{Answers to RQ3:}  Experts evaluated the quality of generated ACs across relevance, correctness, understandability, and coverage, assigning ratings mostly near 4 out of 5 (indicating agreement with quality). Minor correctness issues were primarily due to redundancy in LLM-generated outputs. Overall, experts showed strong enthusiasm for the generated criteria, highlighting instances where the approach captured valuable, non-obvious aspects. They expressed a clear interest in practical adoption, underscoring our approach's real-world applicability and effectiveness.
}

\section{Threats to Validity}
\label{sec:threats}
\textbf{Internal validity.} Results could be distorted by (i) coding errors and (ii) variability in the ``LLM-as-judge'' procedure. To minimise bugs, we mainly adopted the official implementation from the original repositories and applied code reviews of the outputs. For the LLMs' settings, we fixed prompts, temperature = 0, and evaluation heuristics, and split each AC into atomic objectives so that every judge saw the same granularity. Additionally, social-desirability bias~\cite{grimm2010social}, was addressed by ensuring our experts were blind to the system's internals during the interview and by instructing them to justify every decision in writing, a known remedy for this bias. \textbf{Construct validity.} Two complementary families of metrics were used: (a) human-style judgements (Hit/Correct, pairwise preference) and (b) surface statistics (BLEU, ROUGE, Levenshtein, etc). Either family alone is imperfect, e.g., n-gram scores often miss semantic improvements, while LLM judgements can drift. Thus, we collaboratively reported where they diverged. This dual-view mitigates the risk that any single metric misrepresents quality. \textbf{External validity.} We acknowledge the potential limitation of a single industry study, which is always a concern. Our industry study spans only two feature sets of \projName, so wider replications within \company{} and beyond are needed. Such research is rare in RE because few industry or public datasets exist. As all data here is proprietary, we have anonymized a small subset of the data and published every other technical artifact publicly available to aid reproducibility.

\section{Conclusion and Future Work}
\label{sec:conclusion}
In this paper, we introduce \approachName, an automated method for generating acceptance criteria derived from user stories, and demonstrate its efficiency through an industrial project from \company. \approachName~leverages multi-modal requirements processing by integrating advanced Large Language Models (LLMs) with both textual and visual retrieval-augmented generation (RAG) support. We evaluate \approachName~on two samples from \company, focusing on \projName~course structuring for learners and \projName~overall system management for all users, spanning eight stage choices. Based on an interview survey with three experts from \company, we assess the usefulness of \approachName~in terms of correctness, relevance, understandability, and coverage. The experts agree with all four dimensions, indicating that \approachName~achieves high human satisfaction from real-world application perspectives. Furthermore, \approachName~proves beneficial in assisting analysts, whether new to the project or experienced, by reducing manual effort and expanding analytical thinking.

\noindent\textbf{Future work:} We plan to (1) extend the multi-modal RAG pipeline to support additional test artifact generation, as our pipeline can be generalized; (2) validate \approachName~across a broader set of industrial cases to further demonstrate its efficiency and generalization capabilities.

\newpage
\balance
\bibliographystyle{IEEEtran}
\bibliography{paper}

\end{document}